\documentclass{Interspeech}

\usepackage{comment}
\usepackage{footmisc}
\usepackage{color, array, amsthm}
\usepackage{amsmath,amsfonts}
\usepackage{algorithmic}
\usepackage{algorithm}
\usepackage{array}
\usepackage[caption=false,font=normalsize,labelfont=sf,textfont=sf]{subfig}
\usepackage[table]{xcolor}
\usepackage{textcomp}
\usepackage{stfloats}
\usepackage{url}
\usepackage{verbatim}
\usepackage{graphicx}
\usepackage{multicol, multirow}
\usepackage{threeparttable}
\usepackage{tablefootnote}
\usepackage{booktabs}
\usepackage{bbm}
\usepackage[bb=boondox]{mathalfa}
\usepackage{slashed}
\usepackage{scalerel,amssymb}
\usepackage{stackengine}
\usepackage{tipa} 

\interspeechcameraready 

\title{Rethinking Leveraging Pre-Trained Multi-Layer Representations \\for Speaker Verification}

\author{Jin Sob}{Kim}
\author{Hyun Joon}{Park}
\author{Wooseok}{Shin}
\author{Sung Won}{Han\text{$^*$}}

\affiliation[nocounter]{School of Industrial and Management Engineering}{Korea University}{Republic of Korea}
\email{
    \{jinsob,winddori2002,wsshin95,swhan\}@korea.ac.kr
}

\keywords{speaker recognition, speaker verification, speech pre-trained model, fine-tuning efficiency, multi-level features}

\newcolumntype{G}{>{\color[rgb]{0.55,0.55,0.55}\fontsize{7pt}{9pt}\selectfont}l}

\begin{document}

\maketitle

\begin{abstract}
Recent speaker verification studies have achieved notable success by leveraging layer-wise output from pre-trained Transformer models.
However, few have explored the advancements in aggregating these multi-level features beyond the static weighted average.
We present Layer Attentive Pooling (LAP), a novel strategy for aggregating inter-layer representations from pre-trained speech models for speaker verification.
LAP assesses the significance of each layer from multiple perspectives time-dynamically, and employs max pooling instead of averaging.
Additionally, we propose a lightweight backend speaker model comprising LAP and Attentive Statistical Temporal Pooling (ASTP) to extract speaker embeddings from pre-trained model output.
Experiments on the VoxCeleb benchmark reveal that our compact architecture achieves state-of-the-art performance while greatly reducing the training time.
We further analyzed LAP design and its dynamic weighting mechanism for capturing speaker characteristics.\footnote{\texttt{https://github.com/sadPororo/LAP}}
\end{abstract}

\section{Introduction}
\renewcommand{\thefootnote}{\fnsymbol{footnote}}
\footnotetext[1]{Corresponding author} 
\renewcommand{\thefootnote}{\arabic{footnote}} 
Speaker Verification (SV) aims to authenticate an individual's identity based on their unique vocal characteristics.
In recent years, unveiling large labeled datasets \cite{nagrani17_interspeech, chung18b_interspeech} and advancements in deep-learning technologies have led to remarkable improvements in this field.
Considerable efforts have been dedicated to developing sophisticated model architectures \cite{snyder18_xvector, okabe18_attnpool, zhu18_attnpool, snyder2019_etdnn, yu2020_dtdnn, zhang2020_aret, desplanques20_interspeech, heo2024next} and training objectives \cite{wang2018_amsoftmax, deng_2019, deng_2020} to extract distinct speaker representations from acoustic features.

Meanwhile, the pre-training paradigm with Transformer models \cite{schneider19_wav2vec, baevski20_wav2vec2, hsu21_hubert, chen2022_wavlm} has achieved significant success in speech processing.
Attained through speech predictive \cite{oord2018_cpc} or denoising \cite{hsu21_hubert, chen2022_wavlm} modeling, such models offer powerful features for downstream tasks; boosting performance while also leading to faster training convergence.

Various approaches have been explored to utilize pre-trained representations in SV.
For instance, \cite{fan21_explorewv2} and \cite{vaessen22_finetunewv2} fine-tuned wav2vec 2.0 \cite{baevski20_wav2vec2} to extract speaker embeddings directly.
To obtain the speaker vector, the former averaged the output of the last layer, while the latter inserted a constant $cls$ token into the input sequence of the Transformer encoder.
\cite{novoselov23_wav2vec2_tdnn} applied a time-delay neural network (TDNN)-based backend architecture to transform the pre-trained model output into speaker embedding.
Previous studies \cite{chen2022_wavlm, chen22_largescale_for_asv} have proposed utilizing layer-wise outputs from pre-trained models with a powerful backend speaker extractor, ECAPA-TDNN \cite{desplanques20_interspeech}.
To achieve state-of-the-art verification performance, these methods employed a weighted sum of multiple hidden states as input to the speaker model, as introduced in the SUPERB benchmark \cite{yang21_superb}.
These two cases inspired subsequent SV studies to adopt similar strategies to leverage pre-trained models \cite{peng2022_slt, peng2024_finetune_ptm}.
Emphasizing the efficiency of the fine-tuning, \cite{peng2022_slt} designed an attention-based backend module that is lightweight and convolution-free.
Dual-branch ECAPA-TDNN \cite{peng2024_finetune_ptm} was proposed with a multi-level fusion strategy that combines outputs from a pre-trained model and hand-crafted features.

On the other hand, both \cite{peng2022_slt} and \cite{peng2024_finetune_ptm} discussed the underutilization of high-level representations, as shown in \cite{chen2022_wavlm, chen22_largescale_for_asv}, and attempted to exploit features from all levels.
They separately applied a weighted sum on multiple branches, either of key-value flow or of dividing low-high-level layers.
However, we fundamentally question the SUPERB strategy for incorporating layer-wise features to address this issue.
Given the static weights for each layer, Softmax-based aggregation favors low layers, potentially constraining the exploitation of high-level speech attributes \cite{baevski20_wav2vec2, hsu21_hubert} such as phonemes and syllables.

In this paper, we discuss the time-dynamic utilization of multi-layer representations and the effective integration of all layers, from pre-trained models for SV. 
The main contributions of this study can be summarized as follows:
\begin{itemize}
    \item{
        We propose Layer-wise Attentive Pooling (LAP), which applies time-dynamic weighting to multi-layer representations from pre-trained models.
        LAP effectively leverages these representations by addressing the issue of neglecting certain layers in the conventional weighted summation approach.
    }
    \item{
        Aiming for efficient fine-tuning, we introduce a lightweight backend speaker extractor comprising two attentive pooling modules: LAP for layer-wise aggregation and attentive statistics pooling \cite{desplanques20_interspeech} to capture temporal dynamics.
    }
    \item{
        The efficacy of the proposed method was validated experimentally.
        Despite the fewer parameters, our model achieves state-of-the-art results on the VoxCeleb benchmark, including 0.37\% EER on Vox-O and robust performance on Vox-H.
        The impact of the proposed inter-layer aggregation strategy on model performance is verified in further analyses.
    }
\end{itemize}

\section{Extracting Speaker Embedding with Pre-Trained Models}
\begin{figure*}[!t]
  \centering  
  \includegraphics[width=\linewidth]{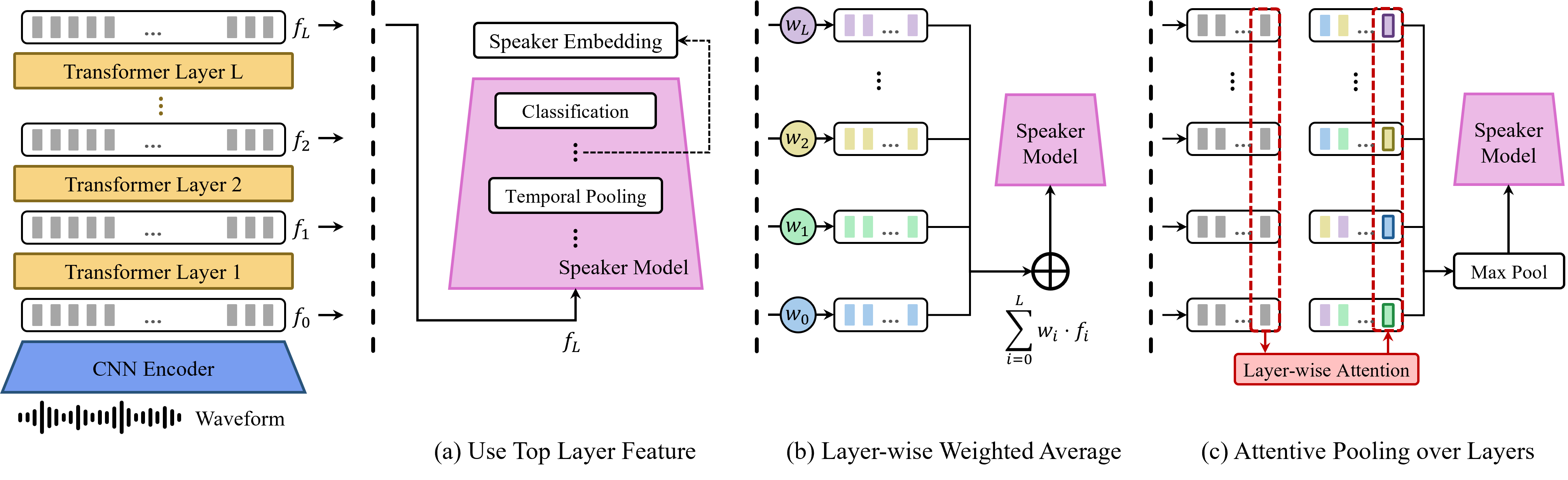}
  \caption{Approaches for leveraging latent features from speech pre-trained networks in speaker verification.}
  \label{fig:leveraging_pretrained_feature}
\end{figure*}

A typical neural-network-based SV system consists of two parts: the frontend, which extracts acoustic features from raw waveform speech, and the speaker model, which processes variable-length features into a fixed-dimensional vector representing the speaker characteristics.
Conventionally, the frontend extracts hand-crafted features such as MFCC and FBank, two-dimensional data, consisting of time (frames) and frequency channels \cite{snyder18_xvector, desplanques20_interspeech, heo2024next}.
Therefore, speaker models have been developed to handle frame-level information processing and temporal pooling effectively to generate robust speaker embeddings.

Recent studies have adopted Transformer-based pre-trained networks to directly process raw waveform speech.
The various approaches for leveraging pre-trained models are illustrated in Figure \ref{fig:leveraging_pretrained_feature}.
Previous studies \cite{fan21_explorewv2, vaessen22_finetunewv2, novoselov23_wav2vec2_tdnn} exploited top-layer representations to extract speaker embeddings, as shown in Figure \ref{fig:leveraging_pretrained_feature}(a).
Figure \ref{fig:leveraging_pretrained_feature}(b) illustrates the weighted summation strategy \cite{yang21_superb}, which aggregates the hidden states from each encoder layer.
The majority of studies \cite{chen2022_wavlm, peng2022_slt, chen22_largescale_for_asv, peng2024_finetune_ptm} have adopted this approach to achieve state-of-the-art performance and demonstrated the importance of utilizing hidden states from encoder layers.
However, as \cite{peng2022_slt, peng2024_finetune_ptm} pointed out, simple weighted summation often overlooks certain layers, leading to the suboptimal utilization of pre-trained representations.
To address this issue, we introduce a novel layer-wise pooling approach that gauges time-dynamic layer weighting, as shown in Figure \ref{fig:leveraging_pretrained_feature}(c).

In addition to the novel layer-pooling strategy, we propose a speaker model composed solely of pooling architectures: LAP and Attentive Statistic Temporal Pooling (ASTP).
Given that the pre-trained models can be fine-tuned to produce more suitable features for downstream tasks, its shallow and lightweight design not only allows efficient training but also promotes the frontend model to derive more robust speaker information.
The following explains the process of extracting the speaker embedding from a raw speech utterance.

\subsection{Pre-Trained Models}
Self-supervised models for large-scale speech training, such as wav2vec 2.0 \cite{schneider19_wav2vec}, HuBERT \cite{hsu21_hubert}, and WavLM \cite{chen2022_wavlm}, process raw waveform inputs directly and provide general speech representations.
As shown on the left-hand side of Figure \ref{fig:leveraging_pretrained_feature}, these models have a common structure, which is a Transformer network above the temporal convolution layers.
We employed WavLM, which is attested to be most suitable for SV \cite{chen2022_wavlm, peng2022_slt, peng2024_finetune_ptm}.
Given a speech utterance  $x \in \mathbb{R}^{\mathbf{T}}$, the multi-layer representations extracted from WavLM can be formulated as:
\begin{equation}
\begin{aligned}
[f_0,f_1, ..., f_L] &= \text{WavLM}(x)
\end{aligned}
\label{eq:wavlm_outputs}
\end{equation}
where $f_{l} \in \mathbb{R}^{C \times T}$ denotes each $l^{\text{th}}$ hidden state; $C$ represents the hidden dimension; and $T$ is for number of frames.

\subsection{Layer Attentive Pooling}
To fully leverage the multi-layered nature of pre-trained models, we aimed to develop a lightweight yet effective method for aggregating more refined layer-wise information.
Accordingly, we adopt an attentive operation \cite{kim2023way, kim2024universal} that combines multi-head projection \cite{vaswani2017_transformer} with the Squeeze--Excitation (SE) \cite{hu2018_seblock}.
Let us begin with projecting $X \in \mathbb{R}^{C \times L \times T}$, which is a stack of $[f_0, ..., f_L]$, with a learnable matrix $W_\text{in} \in \mathbb{R}^{d \times C}$.
\begin{equation}
\begin{aligned}
    x = W_\text{in} \cdot X \hspace{2pt} \in \mathbb{R}^{d \times L \times T}
\end{aligned}
\label{eq:in_projection}
\end{equation}
Next, we pool the statistics of $x$ for the latent dimension $d$ as:
\begin{gather}
    x_{\text{max}} = \text{max}_{d}\text{ }x, \quad  x_{\text{mean}} = 1/d \cdot \Sigma_{d}\text{ }x 
\end{gather}
yielding $x_{\text{max}}, x_{\text{mean}} \in \mathbb{R}^{L \times T}$.\\
Then, we derive layer-wise significance at each time frame as:
\begin{gather}
    \text{SE}(z) = W_{\text{ex}} \cdot \text{ReLU}( W_{\text{sq}} \cdot z), \\
    \alpha = \sigma(\text{SE}(x_{\text{max}}) + \text{SE}(x_{\text{mean}}))
\end{gather}
where $W_{\text{sq}} \in \mathbb{R}^{\gamma \times L}$ and $W_{\text{ex}} \in \mathbb{R}^{L \times \gamma}$ are learnable parameters shared within the statistics, and $\gamma = 0.5L$.
Sigmoid activation $\sigma(\cdot)$ is applied to produce a weight matrix $\alpha \in \mathbb{R}^{L \times T}$.\\
Finally, we scale the layer-wise latent features and then apply max-pooling over layers, resulting in $y \in \mathbb{R}^{d \times T}$.
\begin{equation}
\begin{aligned}
    y = \text{max}_L\text{ }(\alpha \odot x)
\end{aligned}
\label{eq:max_pooling}
\end{equation}
Formally, the series of processes in Equations (\ref{eq:in_projection})--(\ref{eq:max_pooling}) defines LAP given input $X$, and we use $h$ LAP heads as follows:
\begin{gather}
    \text{head}_i = \text{LAP}_i(X) \\
    \text{LAP}(X) = \text{Norm}(\text{Concat}(\text{head}_1, ..., \text{head}_h) \cdot W_\text{out})
\end{gather}
where $W_\text{out} \in \mathbb{R}^{R \times (h \cdot d)}$ is a learnable matrix.
We set $h$ and $d$ to be consistent with those of the pre-trained model, and $R=512$.

\subsection{Attentive Statistic Temporal Pooling}
We adopted channel- and context-dependent statistical pooling \cite{desplanques20_interspeech} to transform frame-level features $x \in \mathbb{R}^{C \times T}$ into utterance-level embedding. 
This technique estimates $\alpha \in \mathbb{R}^{C \times T}$ the importance of each frame given the channel from its channel-wise statistics.
Then, it outputs channel-wise weighted statistics, mean and standard deviation as $\tilde{\mu}=\sum^T_{t}{\alpha_{c,t}x_{c,t}}$ and $\tilde{\sigma} = \sqrt{\sum^T_t{\alpha_{c,t}x_{c,t}^2} - \tilde{\mu}_c^2}$.
The final speaker embedding is produced via linear projection $W \in \mathbb{R}^{R \times 2C}$ and normalization, where $C=512$ and $R=192$.

\section{Experimental Setup}
We conducted experiments on the VoxCeleb datasets \cite{nagrani17_interspeech, chung18b_interspeech}. 
VoxCeleb2 development set was employed for training and the verification performance was evaluated using VoxCeleb1 trial protocols.
During training, we randomly cropped 2s segments from each sample.
The MUSAN \cite{musan2015} and RIR \cite{rir2019} datasets are used to add noise and reverberation.
Speaker augmentation \cite{yamamoto19_interspeech} was applied with speed perturbation factors of 0.9 and 1.1.
We trained our model using the Adam \cite{KingmaB_2015} optimizer with the mini-batch size of 512 and one-cycle scheduling \cite{smith_2019}.
Using a scale of 30, we employed the Additive Angular Margin (AAM) Softmax loss with three sub-centers \cite{deng_2019, deng_2020}.
With some minor modifications, we followed the training process as \cite{chen2022_wavlm, chen22_largescale_for_asv}.
\\
\textbf{Speaker Model Pre-Training}
During this stage, only the speaker model parameters were updated, whereas the frontend remained frozen.
The learning rate was adjusted between $1e^{-5}$ to $1e^{-3}$, using the initial 15\% of the training iterations as warm-up steps.
A weight decay of $5e^{-5}$ was applied to all trainable parameters, including the AAM-Softmax layer.
First, we set the AAM-Softmax margin to 0.0 and then increased it to 0.3 in log-scale for the initial 20 epochs.
Additionally, an Inter-TopK penalty \cite{zhao_2022} is applied to the top-five misclassified centers, with the penalty proportionally scaled alongside the AAM-Softmax margin, reaching up to 0.06.
\\
\textbf{Joint Fine-Tuning}
We unfreeze the frontend model weights and disable speaker augmentation.
The learning rate was gradually reduced to $5e^{-6}$ over 10 epochs, and a weight decay of $1e^{-5}$ was applied throughout the fine-tuning process.
Additionally, we conducted three extra training epochs, following the strategy introduced as large-margin finetuning \cite{qmf2021_lmft}.
For AAM-Softmax, a margin of $0.5$ was used, and no penalty was applied for misclassifications during additional iterations.
\\
\textbf{Evaluation}
The models were assessed using the Equal Error Rate (EER) and minimum Detection Cost Function with $C_{FA} = C_{Miss} = 1$, considering $P_{target} = 0.01$ (DCF$_{1}$) and $0.05$ (DCF$_{5}$).
We used cosine similarity and the adaptive s-norm \cite{matejka17_interspeech} with an imposter cohort size of 600 to score the trials.
As in \cite{qmf2021_lmft, chen2022_wavlm}, a quality-aware score calibration was applied, in which we sampled 30K pairs from the training set.

\begin{table*}[t]
\caption{Performance comparison with state-of-the-art methods on the VoxCeleb1-clean protocol. $\dagger$ denotes the reproduced result, and * is for the estimated value. Lower values are better performance across all metrics, and EER is reported in percentile units (\%).}
\centering
\fontsize{8pt}{9.5pt}\selectfont
\renewcommand{\tabcolsep}{1mm}
    \begin{tabular}{ll l@{}c r@{ }l c ccc c ccc c ccc}
        \toprule
        \multicolumn{2}{l}{ \multirow{2.5}{*}{Frontend} } & \multicolumn{2}{c}{ \multirow{2.5}{*}{Speaker Model} } & \multicolumn{2}{c}{ \multirow{2.5}{*}{Param (M)} } &
            & \multicolumn{3}{c}{Vox-O} && \multicolumn{3}{c}{Vox-E} && \multicolumn{3}{c}{Vox-H} \\
        \cmidrule{8-10} \cmidrule{12-14} \cmidrule{16-18}
            &&&&&&& EER & DCF$_{1}$ & DCF$_{5}$ && EER & DCF$_{1}$ & DCF$_{5}$ && EER & DCF$_{1}$ & DCF$_{5}$ \\
        \midrule
        \multicolumn{2}{l}{ Fbank-80 } & $x$-vector$^{\dagger}$ && \multicolumn{2}{c}{ 4.6 } &
                    & 3.86 & 0.383 & 0.230 &
                    & 3.94 & 0.399 & 0.255 &
                    & 6.55 & 0.512 & 0.360 \\
                 && ECAPA-TDNN {\fontsize{7.5pt}{9pt}\selectfont ($C$=1024)} & \cite{desplanques20_interspeech} & \multicolumn{2}{c}{ 14.7 } &
                    & 0.87 & 0.107 & - &
                    & 1.12 & 0.132 & - &
                    & 2.12 & 0.210 & - \\
                 && NeXt-TDNN {\fontsize{7.5pt}{9pt}\selectfont ($C$=256, $B$=3)} & \cite{heo2024next} & \multicolumn{2}{c}{ 7.1 } &
                    & 0.79 & 0.087 & - &
                    & 1.04 & 0.115 & - &
                    & 1.82 & 0.182 & - \\
        \specialrule{0.4pt}{2.0pt}{1pt}
        \specialrule{0.4pt}{1pt}{2.5pt}
        WavLM & \textsc{Base}   & $x$-vector$^{\dagger}$ && 94.4 & + 6.4 &
                                & 2.25 & 0.373 & 0.225 &
                                & 2.35 & 0.293 & 0.159 &
                                & 4.70 & 0.510 & 0.317 \\
                    && ECAPA-TDNN {\fontsize{7.5pt}{9pt}\selectfont ($C$=512)}$^{\dagger}$ &&& + 8.0 &
                                & 0.85 & 0.119 & 0.062 &
                                & 1.01 & 0.107 & 0.062 &
                                & 1.93 & 0.195 & 0.120 \\
                    && MHFA & \cite{peng2022_slt} && + \textbf{0.4} &
                                & 0.89 & 0.094 & \textbf{0.056} &
                                & 1.09 & 0.117 & 0.068 &
                                & 2.27 & 0.231 & 0.142 \\
                    \rule{0pt}{2.25ex}
                    && \textbf{LAP} {\fontsize{7.5pt}{9pt}\selectfont ($h$=12)} + \textbf{ASTP} & (ours)  && + 1.7 &
                                & \textbf{0.78} & \textbf{0.087} & \textbf{0.056} &
                                & \textbf{0.89} & \textbf{0.095} & \textbf{0.056} &
                                & \textbf{1.72} & \textbf{0.162} & \textbf{0.101} \\
        \specialrule{0.4pt}{2.0pt}{2.5pt}
        & \textsc{Base+} & $x$-vector$^{\dagger}$ && 94.4 & + 6.4 &
                                & 1.82 & 0.411 & 0.193 &
                                & 1.99 & 0.275 & 0.147 &
                                & 4.22 & 0.544 & 0.315 \\
                    && ECAPA-TDNN {\fontsize{7.5pt}{9pt}\selectfont ($C$=512)} & \cite{chen2022_wavlm} && + 8.0$^{*}$ &
                        & 0.84 & -     & -     &
                        & 0.93 & -     & -     &
                        & 1.76 & -     & -     \\
                    && MHFA {\fontsize{7.5pt}{9pt}\selectfont (64 heads)} & \cite{peng2022_slt} && + 2.2 &
                        & \textbf{0.59} & \textbf{0.069} & \textbf{0.041} &
                        & 0.79 & 0.089 & 0.050 &
                        & 1.73 & 0.177 & 0.107 \\
                    && DBE {\fontsize{7.5pt}{9pt}\selectfont ($C$=512)} & \cite{peng2024_finetune_ptm} && + 14.3$^{*}$ &
                        & 0.63 & 0.077 & - &
                        & 0.81 & 0.091 & - &
                        & 1.64 & 0.166 & - \\
                    \rule{0pt}{2.25ex}
                    && \textbf{LAP} {\fontsize{7.5pt}{9pt}\selectfont ($h$=12)} + \textbf{ASTP} & (ours) && + \textbf{1.7} &
                        & 0.61 & 0.091 & 0.050 &
                        & \textbf{0.77} & \textbf{0.087} & \textbf{0.046} &
                        & \textbf{1.49} & \textbf{0.146} & \textbf{0.086} \\
        \specialrule{0.4pt}{2.0pt}{2.5pt}
        & \textsc{Large} & $x$-vector$^{\dagger}$ && 315.5 & + 7.0 &
                                & 1.52 & 0.304 & 0.167 &
                                & 1.87 & 0.264 & 0.138 &
                                & 4.08 & 0.514 & 0.304 \\
                    && ECAPA-TDNN ($C$=512) & \cite{chen2022_wavlm} && + 8.6$^{*}$ &
                        &          0.38 & -     & -     &
                        & \textbf{0.48} & -     & -     &
                        & \textbf{0.99} & -     & -     \\
                    && MHFA {\fontsize{7.5pt}{9pt}\selectfont (64 heads)} & \cite{peng2022_slt} && + \textbf{2.2}\tablefootnote{The open-source model is implemented with 2.37M parameters.} &
                        & 0.49 & 0.091 & 0.045 &
                        & 0.80 & 0.084 & 0.049 &
                        & 1.70 & 0.163 & 0.101 \\
                    \rule{0pt}{2.25ex}
                    && \textbf{LAP} {\fontsize{7.5pt}{9pt}\selectfont ($h$=16)} + \textbf{ASTP} & (ours) && + \textbf{2.3} &
                        & \textbf{0.37} & \textbf{0.059} & \textbf{0.029} &
                        & 0.50 & \textbf{0.055} & \textbf{0.032} &
                        & 1.01 & \textbf{0.099} & \textbf{0.061} \\
        \bottomrule
    \end{tabular}
    \label{tab:main_result}
\end{table*}

\section{Results}
\subsection{VoxCeleb benchmark}
In Table \ref{tab:main_result}, the proposed method is compared with other state-of-the-art verification systems leveraging WavLM.
Except for the proposed method, all other approaches incorporate layer-wise representations using a weighted average strategy \cite{yang21_superb}. 
The proposed method achieved unprecedented performance in the experiments using the \textsc{Base} version. 
Furthermore, it consistently obtained competitive results with \textsc{Base+} and \textsc{Large}, surpassing \textsc{Large}--ECAPA-TDNN \cite{chen2022_wavlm}, which is a widely recognized and powerful approach, on Vox-O.
When compared to MHFA \cite{peng2022_slt}, which has a similar model capacity, the proposed method generally demonstrated superior performance, achieving a 14\%-40\% relative EER improvement on Vox-H.
\subsection{Training Efficiency}
\begin{figure}[t]
  \centering  
  \includegraphics[width=0.95\linewidth]{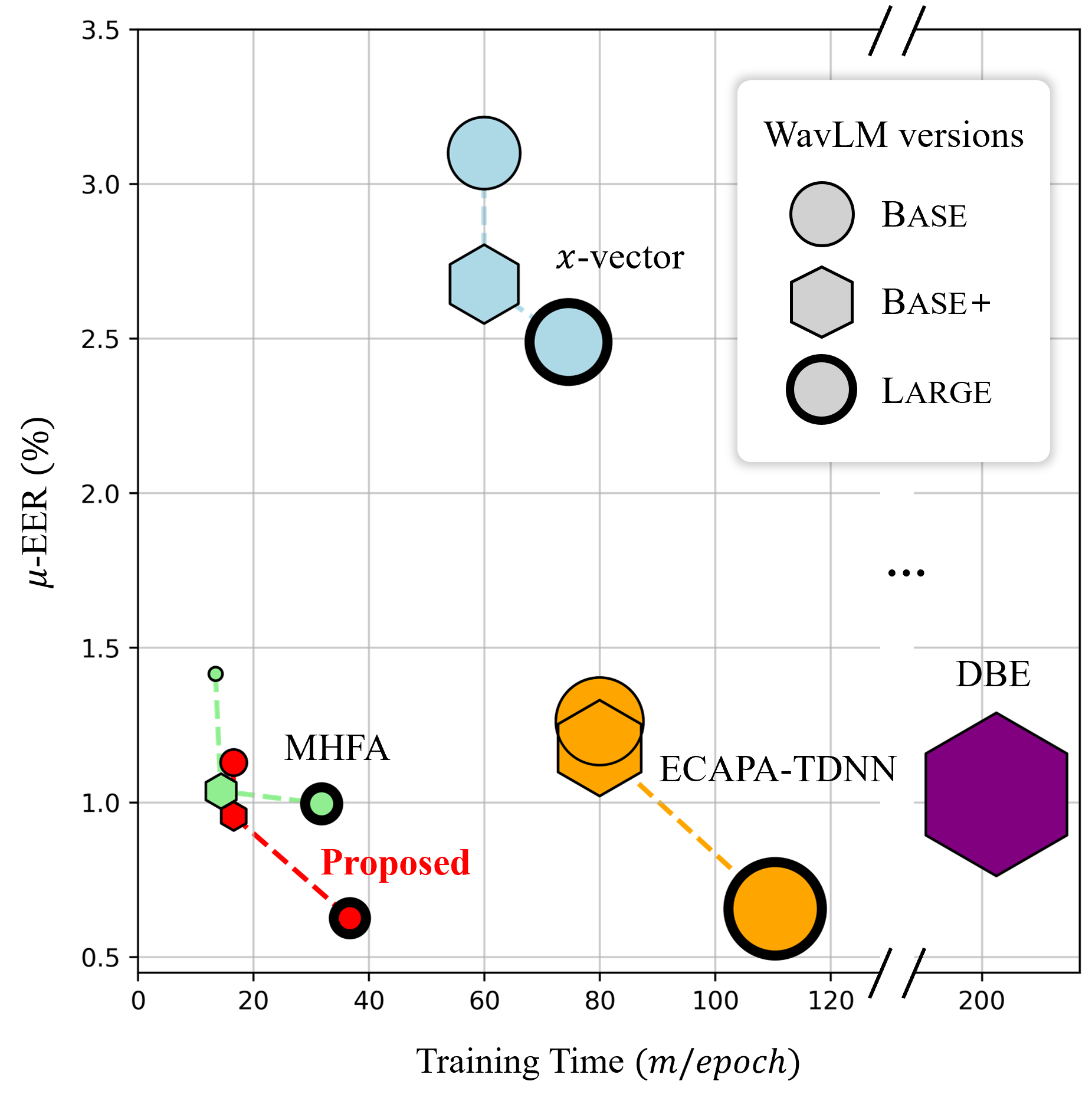}
  \caption{Comparing EER performance and training efficiency.}
  \label{fig:training_efficiency}
\end{figure}

Figure \ref{fig:training_efficiency} shows the performance and training times of the different speaker models. 
$\mu$-EER denotes the average EER across Vox-O, -E, and -H.
The training time was measured for a single epoch during the fine-tuning stage, on four RTX A6000 GPUs.
The marker sizes correspond to the speaker model sizes.
Both MHFA and our proposed method achieved a training time reduction of at least twice compared to TDNN-based models.
This efficiency enhancement stems from the convolution-free and shallow structural design, which significantly reduces the computational overhead.
Notably, our method not only drastically shortens the training time but also outperforms larger models.
In an extreme case, our approach achieved over 10$\times$ faster training while also delivering superior performance compared to DBE \cite{peng2024_finetune_ptm}, which processes dual-branch inputs from FBank and WavLM \textsc{Base+}.

\subsection{Analysis}
\begin{table}[!t]
\caption{Comparison between layer-wise pooling strategies. The frontend size specification is omitted for the \textsc{Base} versions.}
\centering
\fontsize{8pt}{9.5pt}\selectfont
\renewcommand{\tabcolsep}{1mm}
    \begin{tabular}{lr l@{ }l c ccc}
        \toprule
        \multicolumn{2}{l}{\multirow{2.5}{*}{Frontend}} & \multicolumn{2}{l}{\multirow{2.5}{*}{Pooling Strategy}} && \multicolumn{3}{c}{EER (\%)} \\
                \cmidrule{6-8}
                    &&&&& Vox-O & Vox-E & Vox-H \\
        \midrule
        \multicolumn{2}{l}{wav2vec 2.0} & \multicolumn{2}{l}{SUPERB}                        && 2.44 & 2.43 & 4.55 \\
        \cite{baevski20_wav2vec2} && LAP & {\fontsize{7.2pt}{9pt}\selectfont (Softmax-Sum)} && 2.62 & 2.85 & 5.09 \\
                                  &&     & {\fontsize{7.2pt}{9pt}\selectfont (Sigmoid-Max)} && \textbf{1.87} & \textbf{2.08} & \textbf{3.82} \\
        \specialrule{0.4pt}{2.5pt}{1pt}
        \specialrule{0.4pt}{1pt}{3pt}
        \multicolumn{2}{l}{HuBERT}      & \multicolumn{2}{l}{SUPERB}                  && 2.31 & 2.29 & 4.14 \\
        \cite{hsu21_hubert} && LAP & {\fontsize{7.2pt}{9pt}\selectfont (Softmax-Sum)} && 2.62 & 2.96 & 4.98 \\
                            &&     & {\fontsize{7.2pt}{9pt}\selectfont (Sigmoid-Max)} && \textbf{1.80} & \textbf{1.97} & \textbf{3.59} \\
        \specialrule{0.4pt}{2.5pt}{3pt}  
        \multicolumn{2}{r}{\textsc{\quad\quad Large}} & \multicolumn{2}{l}{SUPERB}            && 2.74  & 2.77 & 4.73 \\
                    &                & LAP & {\fontsize{7.2pt}{9pt}\selectfont (Softmax-Sum)} && 2.32  & 2.60 & 4.37 \\
                    &                &     & {\fontsize{7.2pt}{9pt}\selectfont (Sigmoid-Max)} && \textbf{1.73} & \textbf{1.87} & \textbf{3.39} \\
        \specialrule{0.4pt}{2.5pt}{1pt}
        \specialrule{0.4pt}{1pt}{3pt}
        \multicolumn{2}{r}{ WavLM \textsc{ Base+}} & \multicolumn{2}{l}{SUPERB}         && 1.96 & 2.06 & 3.79 \\
        \cite{chen2022_wavlm} && LAP & {\fontsize{7.2pt}{9pt}\selectfont (Softmax-Sum)} && 2.60 & 3.08 & 5.21 \\
                              &&     & {\fontsize{7.2pt}{9pt}\selectfont (Sigmoid-Max)} && \textbf{1.53} & \textbf{1.73} & \textbf{3.24} \\
        \specialrule{0.4pt}{2.5pt}{3pt}  
        \multicolumn{2}{r}{\textsc{\quad\quad Large}} & \multicolumn{2}{l}{SUPERB}            && 1.35  & 1.37 & 2.54 \\
                    &                & LAP & {\fontsize{7.2pt}{9pt}\selectfont (Softmax-Sum)} && 1.21  & 1.26 & 2.32 \\
                    &                &     & {\fontsize{7.2pt}{9pt}\selectfont (Sigmoid-Max)} && \textbf{0.78} & \textbf{1.03} & \textbf{2.10} \\
        \bottomrule
\end{tabular}
\label{tab:ablation}
\end{table}
To verify the efficacy of the proposed layer-wise pooling strategy, we compared SUPERB \cite{yang21_superb} and two types of LAP, applied before ASTP, across popular pre-trained networks, as presented in Table \ref{tab:ablation}.
We report the scores without fine-tuning and without post-processing.
These results demonstrate that the proposed version of LAP excels in utilizing all types of pre-trained models.
A comparison of the two LAPs reveals that the maximum values from the latent features are more effective than the average in terms of capturing speaker-distinct characteristics.
This aligns with the max-pooling design pattern in many classification tasks across different domains, such as image classification.

\begin{figure}[!t]
  \centering  
  \includegraphics[width=\linewidth]{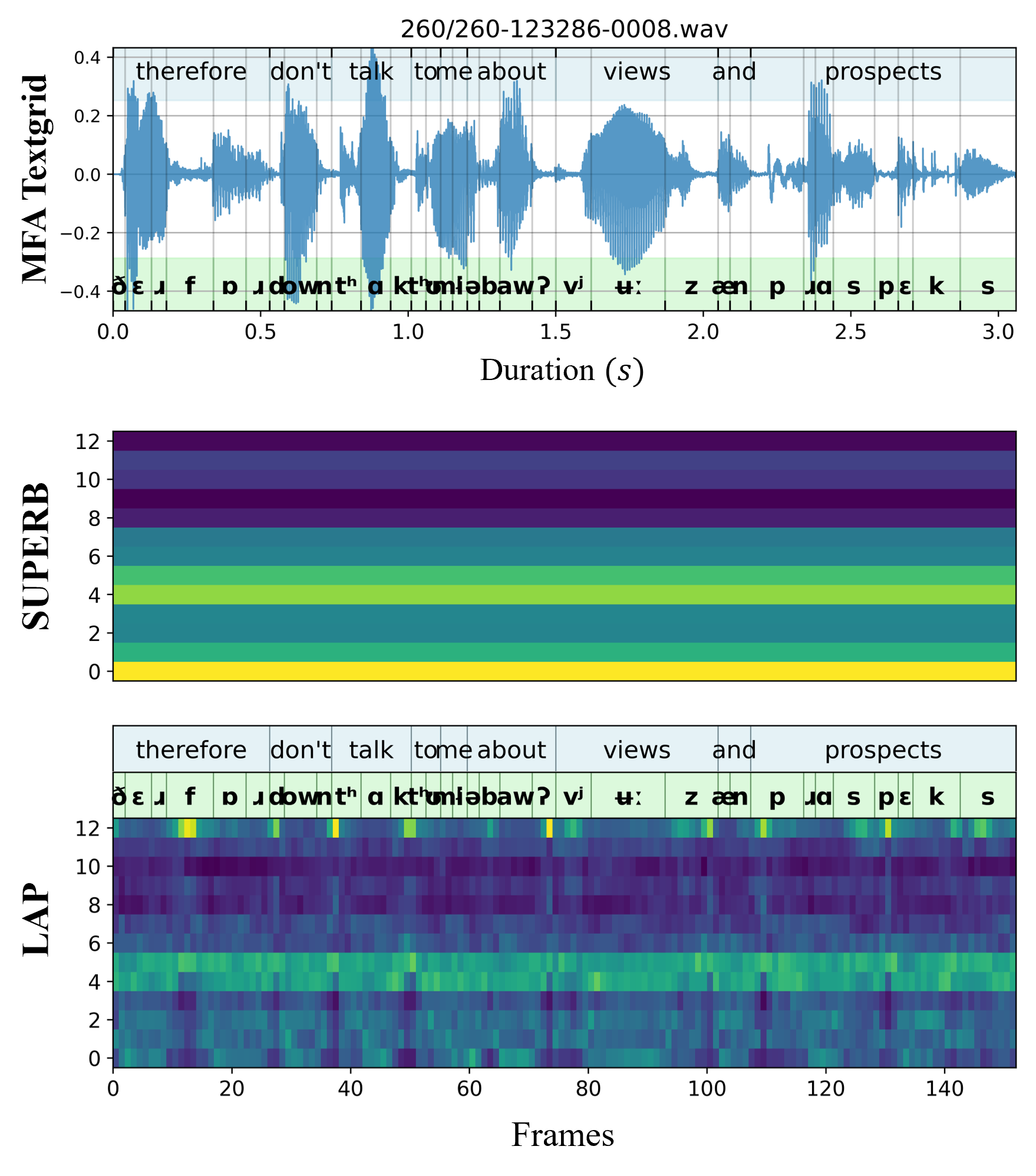}
  \caption{Comparison between the layer aggregation strategies.}
  \label{fig:superb_vs_lap}
\end{figure}

Additionally, we conducted a layer usage analysis to understand how LAP leverages multi-layer features. 
Figure \ref{fig:superb_vs_lap} compares the layer-wise utilization of WavLM \textsc{Base+} between SUPERB and LAP, using a LibriSpeech\footnote{\texttt{https://www.openslr.org/12}} sample.
The visualization reflects how each approach incorporates layers: SUPERB, using a weighted average, is shown by assigned weights, whereas LAP, pooling the undecayed maximum, is represented by the selected count of latent values per layer.
The result highlights the contrast between the time-static and time-dynamic usage of hidden states in WavLM.
SUPERB, regardless of input variation, assigns fixed layer weights, demonstrating a static approach, whereas LAP derives time-variant weights, allowing for adaptive and flexible utilization.
To further investigate, we examined how LAP exploits layer-wise features, using word- and phoneme-level alignments obtained via the Montreal Forced Aligner (MFA)\footnote{\texttt{https://github.com/MontrealCorpusTools}}.
We found that the time-dynamic behavior of LAP aligns with the phonetic and lexical progression of the utterance.
Notably, the top layers respond to the consonants--obstruents, such as $\{\mathbf{f}, \mathbf{t^h}, \mathbf{v^j}, \text{\textbf{\textglotstop}}, \mathbf{z}, \mathbf{p}, \mathbf{s}\}$, contributing to syllable boundary formation in low-level utilization.

\section{Conclusion}
We present LAP, a time-dynamic aggregation method for leveraging multi-level features from pre-trained speech models for speaker verification.
LAP generates layer-wise weights from multi-headed projection, employing SE operation within each head.
Additionally, we propose a lightweight speaker model consisting of LAP and ASTP, pooling architectures for the layer and temporal dimensions respectively.
The proposed model not only achieves state-of-the-art performance on the VoxCeleb benchmark but also demonstrates remarkable training efficiency when utilizing pre-trained models.
Finally, we examined the necessity of the newly introduced max-pooling strategy and demonstrated how LAP dynamically exploits latent features over time by adapting to phoneme-level information.
In future work, we aim to explore combining LAP with other speaker models pushing the limit of performance and further validate its efficacy on diverse speech datasets.

\ifinterspeechfinal
    \section{Acknowledgements}
    This research was supported by the BK21 FOUR funded by the Ministry of Education of Korea and National Research Foundation of Korea. 
    This research was also results of a study on the ``Leaders in INdustry-university Cooperation 3.0" Project, supported by the Ministry of Education and National Research Foundation of Korea.
\else
\fi

\bibliographystyle{IEEEtran}
\bibliography{mybib}

\end{document}